\documentclass[11pt,a4paper]{article}
\usepackage[latin1]{inputenc}
\usepackage{colortbl}
\usepackage{amsmath}
\usepackage{amsfonts}
\usepackage{amssymb}
\usepackage{amsthm}
\usepackage{makeidx}
\usepackage{array}
\usepackage{verbatim}
\usepackage{graphicx}
\usepackage{geometry}
\usepackage{url}

\title{Multi-Linear cryptanalysis in Power Analysis Attacks \\ MLPA} 

\author{
Thomas Roche \\
INRIA EPI Moais / LIG \\
Grenoble University, France \\
Email: thomas.roche@imag.fr
\and
C\'{e}dric Tavernier \\
Communication\&Systems (CS)\\
 Le Plessis Robinson, France \\
Email: cedric.tavernier@c-s.fr
}
\date{16 March 2009}

\begin{document}

\maketitle

\begin{abstract}
Power analysis attacks against embedded secret key cryptosystems are widely studied since the seminal paper of Paul Kocher, Joshua Ja, and Benjamin Jun in 1998 where has been introduced 
the powerful Differential Power Analysis. The strength of DPA is such that it became necessary to develop sound and efficient countermeasures. Nowadays embedded cryptographic primitives 
usually integrate one or several of these countermeasures (e.g. masking techniques, asynchronous designs, balanced dynamic dual-rail gates designs, noise adding, power consumption smoothing, etc. ...).
This document presents a simple, yet interesting,  countermeasure to DPA and HO-DPA attacks, called brutal countermeasure and 
new power analysis attacks using multi-linear approximations (MLPA attacks) based on very recent and still unpublished results of Tavernier et al..
\end{abstract}

\begin{description}
\item[Keywords:] Power Analysis, MLPA, multi-linear cryptanalysis, Reed-Muller codes.
\end{description}

\section{Introduction}      
\label{sec:Introduction}
\paragraph{}
Since the discovery of Differential Power Analysis (DPA) and High Order Differential Power Analysis (HO-DPA) attacks in 1998 (\cite{Kocher98}), the urge to develop resistant hardware
implementations of symmetric ciphers has not ceased. The most popular countermeasures against these devastating attacks have two leaders : the transformed masking methods (initiated by
M.-L. Akkar and C. Giraud in \cite{AkkarGiraud01}) and the duplication method (first proposed by L. Goubin and J. Patarin in \cite{GoubinPatarin99}). When the duplication 
method of rank $n$ has been shown to be vulnerable against a $n$-th order DPA \cite{AkkarGoubin03}, the masking method --- which try to randomize the information leaked from the target 
device --- gave better results in terms of resistance and performances. Thus after several propositions of enhanced DES implementations 
\cite{AkkarGiraud01, AkkarGoubin03, AkkarBevanGoubin04} , the work of Jiqiang Lv and Yongfei 
in 2005 (\cite{JiqiangLvYongfei05}) finally proposed an enhanced version of DES
claimed to be secured against DPA and HO-DPA. To our knowledge, this countermeasure is still holding against those attacks. It uses the unique masking method of \cite{AkkarGoubin03} 
where a new random mask is used for every encryption. Hence,
before each encryption, a set of several custom SBoxes (dependent on the newly generated mask) is generated and stored in RAM. These techniques have the serious drawback of assuming the
SBox generation being done in a secure way (i.e. no information should leak from this operations \cite{AkkarGoubin03}) otherwise it is easy to see that the leaked information would lead to 
HO-DPAs, combining consumptions traces during the SBoxes generation and consumptions traces during the actual encryption. 
From these considerations and the fact that such countermeasures implementations must be thoroughly considered, it is a matter of fact they eventually slowdown the designer of such embedded systems (smartcards, FPGA devices) and then the product's time to market. 
Moreover the resulting implementation, that integrates the additional computations (SBoxes generations), might show itself inefficient in terms of execution time 
from the need of secure computations \cite{AkkarGoubin03}. \\
We present here a brutal way to counter-act Power Analysis attacks. The countermeasure advantages come from its simplicity and how it naturally disable relevant information leakage, 
making it easier to design and implement without assuming that any part of the design is
more secure than another. We will discuss its cost compare to Jiqiang Lv and Yongfei's bounds for DES unique masquing countermeasures \cite{JiqiangLvYongfei05}, thus
isolating some cases where the brutal countermeasure shows itself attractive to designers. Then we introduce a new set of power analysis attacks based on
linear and multi-linear cryptanalysis that will put the first bounds on the brutal countermeasure for DES and AES. Finally we give the current results given by MLPA attacks
on somme simulations and on some real consumption traces (the DPA contest traces found in \url{http://www.dpacontest.org/}).

\section{Preliminaries on embedded symmetric ciphers and Power Analysis attacks}      
\paragraph{}
In this section is first discussed the symmetric cipher design model on which our study has been done and then the way Power Analysis attacks can be applied to those designs.
\subsection{Embedded symmetric cipher design model}
\begin{figure}[h!]
\begin{center}
  \includegraphics[width=15cm,height=4cm]{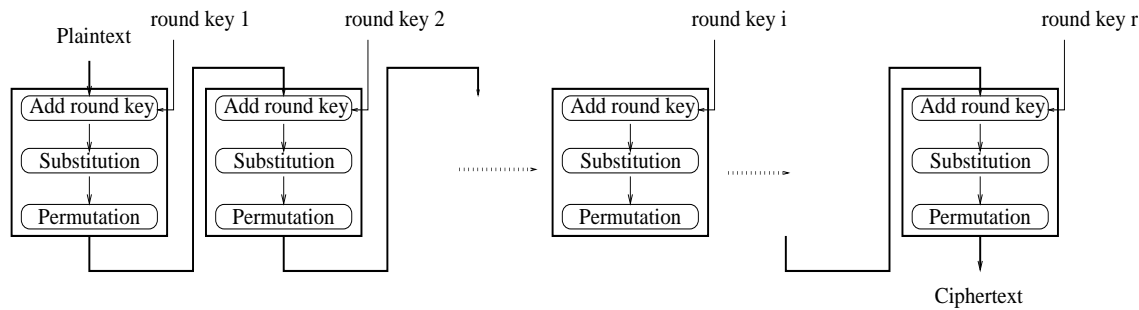}
\end{center}
\caption{Schematic of a Symmetric Cipher}
\label{fig:generic_cipher}
\end{figure}

\paragraph{}
Our study restrict itself to smartcards and FPGA devices that are meant to bore a symmetric cipher implementation. As it is now commonly accepted that hardware implementations of 
symmetric (as well as asymmetric) ciphers achieve at the same time better performances and better security, the development of such devices has tremendously increased in the last few
years. Symmetric cipher hardware implementations can take lots of forms considering the synchronous vs asynchronous designs, the pipelined versions, the implementations 
designed for restricted areas, consumption and/or high throughput. For reasons of clarity, we will describe the studied designs using the common 
shape of symmetric ciphers : Substitution-Permutation Network (SPN) composed in rounds (the key schedule won't be taken in account for our study, we only suppose the round keys
to be available when needed). A symmetric cipher can be represented as on Figure~\ref{fig:generic_cipher} (note that the sub-blocks within a round can be ordered more or less differently).
\paragraph{}
The Permutation part of the cipher, as well as the add round key part are linear functions that can be very efficiently implemented in hardware with simple combinational logic. 
However, the substitution part is usually made of SBoxes, that are highly non-linear functions on 4, 6 (DES) or 8 (AES) bits and are not so easy to implement in combinational 
logic. As a matter of fact, in many designs the SBoxes are stored as lookup tables in memory (RAM or ROM) and accessed when needed in order to save critical logic space. \\
Hence, one way to implement one round of the symmetric cipher is to split it in three clock cycles, the first one dedicated to the add round key function, the second one for
the lookup tables of SBoxes to be accessed and the last one for the diffusion function. Of course each of them can be split again in several clock cycles if needed (In AES for instance,
there can be 8 RAM accesses to the same SBox in one round or just one RAM access if the SBox is duplicated in RAM). \\
Furthermore, when the throughput is more critical than space, it is usually pretty easy to pipeline the executions, in that case it is then mandatory to implement each round 
instead of just one round and a loop counter. \\
To our knowledge Power Analysis attacks on smartcard (ASIC) and FPGA are done on such implementations specifications and they will be the base of our study of PA attacks and countermeasure. 
The knowledge of this high level design (what is computed during each clock cycle) is considered to be known by the attacker, as some probing techniques would give him this information 
anyway.

\subsection{Power Analysis Attacks}
\label{subsec:PAA}
\paragraph{}
Power analysis attack is a dynamic and involved source of research as the development of resistant cryptographic hardware devices is needed. The study of PA attacks and their 
countermeasure has taken a prodigious takeoff since the introduction of the very efficient DPA attacks in 1998. 
\vspace{1ex} \\
\noindent{\textbf{Power consumption in CMOS circuits}
Without going into the depths of CMOS gates power consumption (a simple, yet enough for our need, presentation can be found in \cite{PPA_book} pages 27-60) what we would like to point out here is that 
the power consumption of CMOS circuits is dependent on the data manipulated as transitions from 0 to 1 and 1 to 0 consume significantly more power than 0 to 0 or 1 to 1 transitions through a logical gate.
An attacker observing the overall consumption of a CMOS circuit during two different execution can tell, at a chosen point in time, which execution has led to a greater number
of data changes. What is remarkable to note though is the fact that power consumption of combinational logic (in ASIC or FPGA) at a point within a clock cycle won't give the 
attacker relevant information on the data since one usually assume that the attacker has not a precise enough knowledge of the netlist to be able to predict the glitches occurring throughout the 
logic circuit (see \cite{PPA_book} pages 39-40). Considering this, the power analysis are based on the study of registers and buses power consumption  
since theirs data transitions are synchronized with the clock fronts and don't involve combinational logic. To our knowledge all PA attacks are based on this principle. 
\vspace{1ex} \\
\noindent{\textbf{Hamming distance and Hamming weight models}
When considering the consumption of a bus or register, since the consumption power is significantly higher when a bit value change, the Hamming distance model (HD) says that the power consumption 
is closely related to the Hamming weight of the difference (bit-width Xor) of two successive data values. Note that, of course, absolute values of the measured power traces are not of any use
for the attacker, but relative values with respect to other measurement are relevant. \\
A more simple model, the Hamming weight model (HW), approximate the power consumption directly by the Hamming weight of the manipulated data value. \\
Other models exists, they are basically variants of those models based on some knowledge the attacker might have on the 
targeted hardware design (see \cite{PPA_book} pages 38-43).
\subsubsection{SPA, DPA, HO-DPA}
\label{subsec:DPA}
\paragraph{}
SPA, DPA and HO-DPA attacks are semi-invasive passive attacks introduced in \cite{Kocher98}  by Paul Kocher, Joshua Ja, and Benjamin Jun in 1998. Their semi-invasiveness and passiveness make
them easy to setup, i.e. no need for a complete knowledge of the implementation, timing analysis, and so on. Let us give a rough description of these attacks and introduce some useful notations.
\vspace{1ex} \\
\noindent{\textbf{Simple Power Analysis}
SPA is the simplest way to use Power analysis in order to attack a cryptographic implementation. It requires interpreting the power consumption trace of the cryptographic
function execution. According to \cite{Kocher98}, SPA can be used to break cryptographic implementations in which the execution path depends on the data being processed (e.g. 
conditional branching, comparisons, multipliers, exponentiators, etc. ...). Furthermore the authors consider the prevention of SPA to be fairly simple. 
\vspace{1ex} \\
\noindent{\textbf{Differential Power Analysis}
The efficiency of DPA attacks comes from the fact that instead of studying directly the power consumption over the execution time, it focuses to data-related instructions. By statistical means, DPA
allows the attacker to suppress the measurement noises and bring to light data-dependent operations. Let us borrow the notations of \cite{Kocher98} here : 
\begin{itemize}
  \item $\mathbf{T_i[j]}$ : The $j^{th}$ sample of $T_i$, the $i^{th}$ recorded power trace.
  \item $\mathbf{D(P, B, K_s)}$ : DPA selection function, computes $B$ (Hamming weight of intermediates bits at a fixed point of time), as a function of
  a secret key block $K_s$ and the plaintext $P$ (could also be the ciphertext $C$). In the original DPA from \cite{Kocher98} on DES, $B$ is the Hamming weight 
  of one intermediate bit (i.e. the value of one bit). For now let assume the value of $B$ is 0 or 1.
\end{itemize}
After observing $m$ executions of the cryptographic primitive, recording each power trace $T_{1\cdots m}[1\cdots k]$ ($k$ samples) and the corresponding plaintexts $P_{1\cdots m}$ 
(respectively ciphertexts $C_{1\cdots m}$), the attacker computes the value of $\{B_i\}_{1\cdots m}$ using the selection function $D(P_i, B_i, K_s)$ (for an arbitrary fixed $K_s$). 
The traces are divided in two sets $S_0$ and $S_1$, such that $T_i \in S_0$ iff $B_i = 0$, $T_i \in S_1$ otherwise and the differential trace over the $k$ samples is computed : 
\begin{displaymath}
 \Delta_D[1\cdots k] = \frac{\sum^{m}_{i=1}B_iT_i[1\cdots k]}{\sum^{m}_{i=1}B_i} - \frac{\sum^{m}_{i=1}T_i[1\cdots k]}{\sum^{m}_{i=1}(1-B_i)}
\end{displaymath}
If $K_s$ was a wrong guess, then the values $\{B_i\}_{1\cdots m}$ are not related to the manipulated data and then, when the number of tests increases ($m \rightarrow \infty$), 
the differential trace tends to a flat trace ($\forall j = 1\cdots k, \Delta_D[j] \rightarrow 0$). On an other hand, if $K_s$ was a right guess, the value $\{B_i\}_{1\cdots m}$ 
are correct and the differential trace is related to the power consumption that coincide with the value of $B$. Furthermore, the value of other bits, the measurement noises, 
being not considered by $D$, will less affect the differential trace as the number of tests increases. Hence, the differential trace will bore spikes on samples where the 
manipulated data is correlated with $D$ when $m$ increases. 
\vspace{1ex} \\
\noindent{\textbf{Remarks}
Other methods have been developed to evaluate more or less precisely correlations between the power consumption traces and selections functions,the interested reader can refer to 
the work of E. Brier, C. Clavier and F. Olivier in \cite{BrierClavierOlivier2004} that uses the Pearson coefficient (CPA) and the maximum likelihood method of R. B\'{e}van and E. Knudsen 
\cite{BevanKnudsen2002}. Moreover, when our description details single-bit DPA ($B$ represent a single bit), more complicated selection function can be used where $B$ can
take more than two values (Hamming weight of an intermediate data value), those kind of attacks (DPA multi-bits) have been gathered under the name Partitioning 
Power Analysis (PPA) by Thanh-Ha Le, Jessy Cl\'{e}di\`{e}re, C\'{e}cile Canovas, Bruno Robisson, Christine Servi\`{e}re et Jean-Louis Lacoume in \cite{LeClediereCanovasRobissonServiereLacoume2006}
\vspace{1ex} \\
\noindent{\textbf{High-order DPA}
In a $n$-order DPA, a combination of $n$ points in the data path is involved in the selection function, i.e. for each power trace, $n$ samples will be differentiated in the same
differential trace $\Delta_D(1\cdots n)$.

\subsubsection{Countermeasures}
\label{subsec:old_countermeasure}
\paragraph{}
As introduced in the section~\ref{sec:Introduction}, the unique masking techniques uses random data for every encryption function call in order to randomize the power consumption. Hence, the
additive masking consists in manipulating data that have been xored with a random value (the mask) and follow the mask value throughout the cipher execution such that it can removed 
when needed (at the end of a round, a set of rounds or even at the end of cipher execution). Even though following the additive mask value is pretty easy when considering linear functions,
it show itself tricky when considering highly non-linear function such as SBoxes. Hence, the proposed masking techniques \cite{AkkarGiraud01, AkkarGoubin03, AkkarBevanGoubin04}, uses generations
of custom SBoxes related to the current masks such that the custom SBoxes make it possible to easily follow the mask values. The custom SBoxes are then stored in RAM, the original versions 
of the SBoxes can be stored in RAM or ROM. In \cite{JiqiangLvYongfei05}, the authors proved that three 32-bit random masks and six custom SBoxes are the minimal
cost for a secure DES implementation masking all the outputs of the SBoxes of the sixteen rounds. 

\section{A brutal countermeasure}      
\label{sec:brutalcontermeasure}
\paragraph{}
As has been detailed in the previous section, the power analysis attacks are based on the study of registers or buses power consumptions, as the transitions from one data value to another inside 
them are done at a precise time of the clock cycle and then allows to precisely determine the consumption of such a transition. This consumption being assumed to be closely related to the 
Hamming weight of the manipulated data (straightforwardly in the HW model or on the difference of two successive data in the HD model). DPA attacks work assuming the attacker can predict 
the value of one bit (or of a set bits) actually manipulated by a register or bus as a function of the known input (or output) bits of the cryptographic primitive and few key bits. In practice 
there should not be more than 32 key bits involved \cite{AkkarGoubin03, AkkarBevanGoubin04, JiqiangLvYongfei05} otherwise the attack couldn't be achieved considering the cost in memory 
and acquisition time.  \\
From the above considerations, a straightforward way to disable such Power Analysis attacks is to suppress the use of registers and buses until every bit stored in registers or going 
through the buses are either independent on the secret key or dependent on more than 32 bits of the secret key (i.e. before a certain number of rounds).
\subsection{countermeasure setup and drawbacks}
\paragraph{}
Depending on the target symmetric cipher's diffusion functions, one can fix the number of rounds that must be executed during one clock cycle (i.e. between 
two registers or two access to a bus). Let us consider the two most popular symmetric ciphers : The Data Encryption Standard (DES) and its successor the Advanced Encryption Standard (AES).
The brutal countermeasure for DES would be to compute the first three rounds by pure combinational logic in one clock cycle and, by symmetry, the same thing should be done for the 
last three rounds. For AES, since its diffusion function is more efficient, the first round should be done in one clock cycle as well as the two last rounds (since the last round of AES does not 
contain the diffusion \texttt{MixColumn}). Let us call these incompressible blocks the "glued blocks".\\
The obvious drawback of this countermeasure is that it makes it mandatory to implement the SBoxes in combinational logic (using LUT implementation for instance). Furthermore, on a
pipelined implementation, it would limit the overall throughput (since it forbids to divide the first and last blocks of logic in several clock cycles). \\
The advantages of the countermeasure being its very simplicity to implement (no need for additional functions) and the fact that it does not base itself on a secure pre-computation.\\
It seems important to note here that this countermeasure is not compatible with the unique masking methods since those methods, as seen in section~\ref{subsec:old_countermeasure}, need to 
generate mask-dependent SBoxes at runtime.
\vspace{1ex} \\
\noindent{\textbf{Drawback bypass}
In some cases it is possible to go around the pipeline drawback. When the area is not critical, it is possible to put several glued blocks in parallel
monitored by a slower clock (generated by a pll component for instance) and connect them to the original rounds implementation that runs at a faster clock cycle. This solution would keep a 
high throughput even with the countermeasure. \\
Let us also note that the AES SBox have a very efficient implementation in terms speed and area using the multiplicative inverse function in $GF(2^8)$ (\cite{HodjatVerbauwhede04}). 
\paragraph{}
Finally this countermeasure may be attractive to designers that have a large combinational logic space and give priority to strong security, even though the cost in area is outrageous. 

\section{(M)LPA Attack description and complexity}
\paragraph{}
In this section is introduced Linear Power Analysis and Multy-Linear Power analysis attacks. Those attacks correspond strictly to Linear (\cite{Matsui93}) and Multy-Linear cryptanalysis 
(\cite{KaliskiRobshaw94}) in the side-channel world. We are first going to introduce some useful notations for the study of linear approximations. Then we will introduce the idea of LPA and MLPA 
before describing the attacks algorithms and complexity. Finally we will discuss its practical setup. 
\subsection{Linear approximations of a symmetric cipher}
\label{subsec:linearapprox}
\paragraph{}
Linear cryptanalysis has been introduced by Matsui in 1993 (\cite{Matsui93}), since then it has become one of the most important base of the study of block cipher security. Nowadays new block ciphers 
must prove some inherent resistance against linear cryptanalysis. Let us remark that many cryptanalysis methods are based on this fundamental discovery, among others, the multi-linear cryptanalysis 
\cite{KaliskiRobshaw94, BiryukovDeCannireQuisquater04} will be particularly interesting here.
\vspace{1ex} \\
\noindent{\textbf{linear cryptanalysis}
A linear approximation is defined as a combinations of ciphertext bits as a linear function of plaintext and key bits. \\
Let us denote $|K|$, $|P|$, $|C|$ respectively the bit-lengths of key, plaintext and ciphertext. Let us consider a vector $\Pi$ of length $|P|$, $\kappa$ of length $|K|$ and $\Gamma$ of length $|C|$ and a bit 
$b$. $\Pi$, $\kappa$, $\Gamma$ and $b$ define a linear approximation of bias $\epsilon$ over the symmetric cipher if and only if :
\begin{equation}
  \Pr_{P,K}(<P,\Pi> \oplus <K,\kappa> \oplus b = <C(P,K), \Gamma>) \geq 1/2 + \epsilon
\end{equation}
Given such a linear equation, Matsui showed that a high probability of success to recover the involved key bits in the equation using linear cryptanalysis would require a data-complexity 
(i.e. number of plaintext-ciphertext pairs) of $N=1/\epsilon^2$.
\vspace{1ex} \\
\noindent{\textbf{Multi-linear cryptanalysis}
It was shown in \cite{BiryukovDeCannireQuisquater04} that instead of using a single linear approximation, the use of several linear approximations involving the same key bits would significantly 
improve the performances of the attack. As a matter of fact, given $n$ linearly independent approximations of respective bias $\epsilon_j, j=1,\cdots, n$ the data-complexity of the attack would be 
reduced to 
\begin{displaymath}
  N \approx 1/\sum^n_{j=1}\epsilon^2_j
\end{displaymath}
In a very recent --- yet to be published --- paper, Tavernier et al. (\cite{TavernierToBePublished}, 
studied the problem of finding all the linear approximations with a given bias of a given Boolean function. The authors 
showed the equivalence between the problem of finding linear approximations for a fixed output mask ($\Gamma$ fixed) and a list decoding problem in the first order Reed-Muller code.
They were then able to find good linear approximations up to 8 rounds of DES and thus, based on results of \cite{GerardTillich07}, break a reduced version of the cipher with low 
data-complexity ($2^{21}$ plaintext-ciphertext pairs).
\subsection{Introduction to (M)LPA}
\label{subsec:IntroMLPA}
\paragraph{}
As mentioned above, (M)LPA implies the use of linear approximations to attack a symmetric cipher hardware implementation by power analysis. We will introduce two different ways to 
use linear approximations by an attacker, the later will be the so called (M)LPA attack. Let us denote $H(u)$ the Hamming weight function of a vector of bits $u$.
\vspace{1ex} \\
\noindent{\textbf{A first approach : a classical approach}}
A very straightforward approach would be to attack by DPA, CPA or PPA using a linear approximation as base of the selection function. This will render the attack's selection function dependent on 
the approximation bias $\epsilon$ and thus increase the data complexity. The advantage of such an attack will be to find linear approximation 
that involve few bits of the key (less than $32$ in practice) when evaluating data values in registers or going through buses that are strictly dependent on more than 
$32$ key bits from the point of view of the cipher function. Hence it would allow to attack a cipher implementation where the unique masking technique or
the brutal countermeasure are used only for the data bits that dependent on less than $32$ key bits. \\
For instance let us consider the mono-bit DPA attack presented by Kocher in \cite{Kocher98}. Using the notations introduced in section~\ref{subsec:DPA}, 
let us denote by $m$ the complexity of the attack if the selection function ($D(P, b, K)$) is not probabilistic (classic DPA) and $M$ the one when the selection function ($D_{\epsilon}(P, b, K)$) 
is probabilistic (meaning that $D_{\epsilon}$ has probability $1/2+\epsilon$ to be right). The $k$-sample differential trace $\Delta_D[1\cdots k]$ is then : 
\begin{displaymath}
 \Delta_D[1\cdots k] \approx 2 \left( \frac{\sum^{M}_{i=1}D_{\epsilon}(P_i, b, K)T_i[j]}{\sum^{M}_{i=1}D_{\epsilon}(P_i, b, K)} - \frac{\sum^{M}_{i=1}T_i[j]}{M}\right)
\end{displaymath}
It is easy to see that when the key guess is wrong, the probabilistic section function is not correlated to the manipulated data (as the old selection function) and 
the differential trace will tend to a flat trace when when $M \longrightarrow \infty$. Let us consider now that the key guess is right. Since $D_{\epsilon}$ is right with a probability $p = 1/2+\epsilon$, 
let us denote $D_{true}$ the cases where the selection function is right and $D_{false}$ otherwise. Then, after re-indexing the plaintexts and traces, we have 
\begin{displaymath}
 \begin{array}{lcl}
  \Delta_D[1\cdots k] & \approx & 2 \left(\frac{ \sum^{2\epsilon M}_{i=1}D_{true}(P_i, b, K)T_i[j]}{\sum^{M}_{i=1}D_{\epsilon}(P_i, b, K)} + \frac{\sum^{(1/2+\epsilon)M}_{i=2\epsilon M+1}D_{true}(P_i, b, K)T_i[j]}{\sum^{M}_{i=1}D_{\epsilon}(P_i, b, K)} \right. \\ 
		      &         & \left. + \frac{\sum^{M}_{i=(1/2+\epsilon)M + 1}D_{false}(P_i, b, K)T_i[j]}{\sum^{M}_{i=1}D_{\epsilon}(P_i, b, K)} - \frac{\sum^{M}_{i=1}T_i[j]}{M}\right) \\
                      & \approx & 2 \left(\frac{ \sum^{2\epsilon M}_{i=1}D_{true}(P_i, b, K)T_i[j]}{\sum^{M}_{i=1}D_{\epsilon}(P_i, b, K)} + \frac{\sum^{M}_{i=2\epsilon M+1}\widetilde{D}(P_i, b, K)T_i[j]}{\sum^{M}_{i=1}D_{\epsilon}(P_i, b, K)} - \frac{\sum^{M}_{i=1}T_i[j]}{M}\right)  
 \end{array}
\end{displaymath}
where $\widetilde{D}$ is an uncorrelated selection function (it has 1 chance over 2 to be wrong) and then will tend to a flat trace when $M \longrightarrow \infty$. 
Finally, the data complexity of the attack is such that $2\epsilon M \geq m$, in other words, the complexity of the attack increase by a factor $1/(2\epsilon)$ as the 
selection function has a bias $\epsilon$.  
\vspace{1ex} \\
\noindent{\textbf{Remark 1}} Let us note here that the term $\sum^{M}_{i=1}D_{\epsilon}(P_i, b, K) \approx M/2$ in the above equation will crush the potential spikes amplitude 
and in practice, $\epsilon$ shouldn't have to be very small for a data-complexity to be unreachable in practice. The measurement acquisition time cannot be neglected in 
Power Analysis attacks. \\
\noindent{\textbf{Remark 2}} The attack described above can be easily extended to multi-linear approximation attack.
\vspace{1ex} \\
\noindent{\textbf{Second approach : a HD and HW models approach}} 
An interesting way to use linear approximations would be to directly approximate the Hamming weight of a register since this is the quantity which
is the most correlated to what is being measured.
Thanks to the work of Tavernier et al. (in \cite{TavernierToBePublished}), it is possible to find linear approximations of $<H(C(P,K)), \Gamma_H>$ with 
any chosen vector $\Gamma_H$ ($\Gamma_H$ is a vector of length $log_2(|C|)$, with respect to the notations of section~\ref{subsec:linearapprox}). \\
If we assume that the actual value of the measurement samples $T_i[j]$ is closely related to the value of the hamming weight of the data 
manipulated (for the HW model) or the difference between two successive data manipulated (for the HD model), 
then the use of linear approximations on the hamming weight value of a register (or a bus) would lead to very efficient attacks (a discussion on this assumption is 
given in the later section~\ref{subsubsec:practicalsetup}). 
This important remark is the origin of the new MLPA attacks that should prove themselves much more dangerous than the previous DPA-like approach.

\subsection{The MLPA attack}
\paragraph{}
As introduced in the previous section, the LPA attack is based on the HW and HD models. If we assume that these models are relevant, then multi-linear approximations
can be used in all their strength. As presented in \cite{GerardTillich07, TavernierToBePublished} in the context of classical multi-linear cryptanalysis, one can consider the recovering 
of some key-bits as the decoding problem of a code whose length is equal to the number of available linear relations and over a memoryless channel whose capacity 
depends on the respective biases of the linear approximations. 
Let us consider a set of $n$ linear relations of biases $\epsilon_l, l=1,\cdots, n$ with a form as follow : 
\begin{equation} \label{equa:linearapprox_H}
  <P,\Pi_l> \oplus <H(C(P,K)), \Gamma_{H_l}> \oplus b_l = <K,\kappa_l>
\end{equation}
where the set of vectors $\kappa_l, l=1\cdots n$ are such that a limited number $k$ of key bits are involved in the equations (in practice less than 32 bits) and form a matrix of rank 
$k$ \\
The idea is to reconstruct a code word $y$ of length $2^k$ from a noisy and erased codeword $\tilde{y}$ wich is enough close to $y$, to be able to decode it in the first Reed-Muller code. 

\subsubsection{Attack algorithm}
\label{subsubsec:MLPAalgorithm}
\paragraph{}
After observing $N$ encryptions and selecting the sample $j$ in each traces $T_i, i=1\dots N$ where the target intermediate data bits are manipulated, the attack will proceed as follows :
\begin{enumerate}
\item For each linear approximation and each "plaintext-$T_I[j]$" pair (for the HD model it would be, for each "plaintext pair-$T_i[j]$ pair")
compute the predicted value of $<K,\kappa_l>_i$ using the right member of the equation~\ref{equa:linearapprox_H}
(which would be "$<P_i,\Pi_l> \oplus <T_i[j], \Gamma_{H_l}> \oplus b_l$" since $T_i[j]$ is considered as corresponding to $H(C(P_i,K))$).
\item For each linear approximation, separate the traces into two sets $S_0^l$ and $S_1^l$ for which $<K,\kappa_l>$ has been evaluated to $0$ and $1$ respectively. 
\item Construct the noisy and erased codeword $\tilde{y}$  such that the value of $\tilde{y}$ at position $x_l = \kappa_l$ ($\kappa_l$ is seen here as its value in $GF(2^k)$) is 
$\tilde{y}(x_l) = (\#\{S_0^l\}-\#\{S_1^l\}) ln(\frac{1/2 - \epsilon_l}{1/2 + \epsilon_l}) $. The position were no linear approximation is defined will be put to zero thus considering it 
as an erasure position. 
\item Decode $\tilde{y}$ in the first order Reed-Muller code, i.e. the most probable codeword $y$ is the one that maximise the inner product $\sum_{x\in \{0,1\}^t}(-1)^{y(x)}\tilde{y}(x)$. 
The Fast Fourrier Transform would do the trick in a time complexity $O(k2^k)$ and data complexity $O(2^k)$.
\end{enumerate}
For details of Reed-Muller decoding efficiency in a gaussian and erasure channel, the interested reader should refer to the results of I. Dumer-R. Krichevskiy in 
\cite{DumerKrichevskiy00}. 

\subsubsection{Practical setup} 
\label{subsubsec:practicalsetup}
\paragraph{}               
The attack presented above may seem completely unrealistic since it uses directly the value measured as Hamming weight of the data manipulated, which contradict 
subsequently the remark done in section~\ref{subsec:PAA} on the use of absolute measurement values. Two practical setup seem possible to bypass this :
\begin{itemize}
  \item First of all, let us assume that the targeted device can be run with chosen plaintexts. Under this hypothesis it is possible
        to attack by re-initializing the registers before each encryption (reseting the register would be to run a set of fixed plaintexts until the device is in 
	the same state before each encryption). Therefore, using simple pre-testing on the board, it would be possible to relate the consumption traces to 
	the targeted quantities as following a Gaussian law.
  \item For a more practical attack, assuming that we have access to a twin device where we can put arbitrary chosen keys, it
        would be possible to run the algorithm that search linear approximations directly on the twin device as a pre-processing phase of our attack. 
	As the algorithm is run on a Boolean function 
	as a black box, using the consumption measurement as output value of our Boolean function might render the attack even more efficient than in the model 
	presented above. Further more, it is then possible to mount unknown cipher attacks since no knowledge of the symmetric cipher is needed except for its SPN structure (the hardware 
	device is seen as a black box from which the consumption leakage are the outputs).
        
\end{itemize}

\subsubsection{Results}
\label{subsec:results}
\paragraph{}
In this section are presented the results obtained using the above described attacks on the DES and AES cipher. There are two sets of results, the first ones are called 
simulations and can be seen as the validation of our attack in theoretical model. The second set of experiment have been done on real power traces, and validate the 
practical feasibility of the attack. Table~\ref{table:result_MLPA_simu} and Table~\ref{table:result_MLPA_contest} summarize some of the results, in these tables, 
"\# linear equ." refers to the total number of linear approximations found for the attack, not all of them have been useful, "\# Plaintext" or "\# Traces" refers to the 
data complexity of the attack and "$\Pr$(Success)" refers to the probability of success of the attack in simulation.
\vspace{1ex} \\
\noindent{\textbf{Attack simulations}
The algorithm descried in section~\ref{subsubsec:MLPAalgorithm} has been simulated on the DES and AES cipher. By the means of Tavernier et al.'s work on finding linear approximations,
up to three rounds can be approximated with good enough biases for the hamming weight of an intermediate data value. Hence the figures of Table~\ref{table:result_MLPA_simu} summarize
our results (with respect to HW and HD model). They show that a glued block of three rounds for a DES version of 
the brutal countermeasure wouldn't be enough. The simulation has been done considering that the cipher implementation leakage gives the hamming weight of the targeted data. Hence,
in the HW model, the linear approximations evaluate the hamming weight of the round register (assuming that their is a register after a glued block of 1, 2 or 3 rounds), in
the HD model, the linear approximations evaluate the hamming weight of the differences of the round register between two execution (two different plaintexts). Let us note here that
in a chosen plaintext attack, the HW model results correspond to an HD model.

\begin{table}[!h]
\begin{center}
\tiny{
\begin{tabular}{|c|c|c|c|c|c|c|}
\hline
Cipher & Model & rounds & \# linear equ. & \# key bits & \# Plaintexts & $\Pr$(Success) \\
\hline
\hline
DES    & HW    & 1      &  349           &  30         & $2^{10}$   & 0.79           \\
DES    & HW    & 1      &  349           &  48         & $2^{12}$   & 0.99           \\
\hline
DES    & HW    & 2      &  728           &   6         & $2^{9}$    & 0.97           \\
DES    & HW    & 2      &  728           &   48        & $2^{12}$   & 0.95           \\
\hline
DES    & HW    & 3      &  164           &   12        & $2^{17}$   & 0.96           \\
DES    & HW    & 3      &  164           &   27        & $2^{20}$   & 0.99           \\
\hline
DES    & HD    & 2      &  27            &  16         & $2^{14}$   & 0.71           \\
DES    & HD    & 2      &  27            &  16         & $2^{16}$   & 0.99           \\
\hline
AES    & HW    & Last   &  1410          &  128        & $2^{10}$   & 0.80           \\
AES    & HW    & Last   &  1410          &  128        & $2^{11}$   & 0.99           \\
\hline
\end{tabular}
\caption{Simulation Results}
\label{table:result_MLPA_simu}
}
\end{center}
\end{table}
It is important to note here that no linear approximation have been found for the first round in HD model, as if no information would leak from the hamming weight of the 
data manipulated. The attack on AES has been done on the last round since it does not contains the \texttt{MixColumn} diffusion function.
\vspace{1ex} \\
\noindent{\textbf{Attack on DPA-contest traces}
Thanks to the DPA contest, power consumptions traces are freely available. Unable to obtain and setup a hardware device ourselves, these online available traces allowed us to try our attack on real 
power traces and then prove the feasibility in a real setup of the attack. The attack has been launched on the contest traces (\texttt{secmatv1\_2006\_04\_0809})
that yield about 80000 power consumption traces. The linear approximations evaluate the hamming weight of the difference of data stored in the implementation register ($LR$)
(see \cite{guilleyHoogvorstPascalet07} for more details on the DES implementation), the attack description and setup can be found in Annexe of this document.
\begin{table}[!h]
\begin{center}
\footnotesize{
\begin{tabular}{|c|c|c|c|c|c|c|}
\hline
Cipher & rounds & \# linear equ. & \# key bits & \# traces \\
\hline
\hline
DES    & 1      & 84            &  $\sim$20   &   1000    \\
DES    & 1      & 84            &    45       &   20000   \\
\hline
DES    & 2      & 163           &  $\sim$10   &   1000    \\
DES    & 2      & 163           &    47       &   36000   \\
\hline
\end{tabular}
\caption{Attack on DPA-contest traces Results}
\label{table:result_MLPA_contest}
}
\end{center}
\end{table}
\section{Conclusion and future work}
\paragraph{}
The results shown in section~\ref{subsec:results} prove the feasibility of the MLPA attacks, it is our belief that this set of attacks is a starting point of 
new results on power analysis attacks on embedded symmetric ciphers. Hence the next steps will be of two kinds :
\begin{itemize}
\item The research of better linear approximations in term of bias and which can approximate more rounds of the symmetric cipher. This implies a
complexity in time that we did not have for the redaction of this document. 
\item The experimentation on an unknown cipher implementation with research of linear approximation directly on the board. This attack may lead
to very efficient attacks since it directly approximate the leakage function without using any consumption model. 
\end{itemize}

\bibliographystyle{plain} 
\bibliography{paper}

\section*{Annexe : The attack on DPA-context traces setup}
\paragraph{}
This annexe describe an MLPA attack on power traces found on the dpa-contest website : \url{http://www.dpacontest.org/}. The traces used for our attack are stored 
under the name : \texttt{secmatv1\_2006\_04\_0809}, there is 81089 power traces that have been measured from a straightforward DES implementation detailed in \cite{guilleyHoogvorstPascalet07}. 
\paragraph{}
The implementation is described in the figure~\ref{fig:DES_dpacontest} (from \cite{guilleyHoogvorstPascalet07}). Let us denote $H(X)$ the Hamming weight function, 
$IP(X)$ the initial permutation of DES cipher
and $DES_n(X,K)$ the first $n$ rounds of the DES encryption on a $64$-bits vector $X$ and a $(n\times 64)$-bits $K$. The power measurement samples we are interested in are the ones 
corresponding to the load of the register LR, after round 1 and 2. According to the Hamming Distance model, they should correspond to $H(IP(X) \texttt{ XOR } DES_1(X,K))$ (noted  $C_1(X,K)$) 
and $H(DES_1(X,K) \texttt{ XOR } DES_2(X,K))$ (noted $C_2(X,K)$) respectively.
\begin{figure}[h!]
\begin{center}
  \includegraphics[width=7cm,height=4cm]{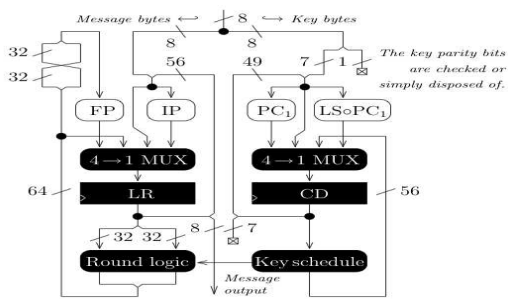}
\end{center}
\caption{Schematic of DES implementation}
\label{fig:DES_dpacontest}
\end{figure}
The sample indexes were found by just simulating a DPA attack on the first round and on the second round (using the first round key). It is our believe that 
these informations could have been found by an attacker using simple timing measurement, anyways it is a hypothesis of the MLPA attack that these informations are known. Hence, the load of 
register LR after the first round (respectively the second round) was found to be corresponding to the $5749$th (respectively the $6374$th) sample of the power traces. 
\paragraph{}
Linear approximations have been generated corresponding to $<C_i(P,K),\Gamma_H>$, $i\in\{1,2\}$. Only the ones where $\Gamma_H$ equals to $0x10$ or $0x20$ were kept. The Table~\ref{table:equations}
give an example of 11 of these approximations for the second round ($C_2$). Over these 11 equation, only 6 key bits are involved ($K[j]$ is the $j$th bit of the secret key). The last thing we 
now have to do in order to apply the MLPA algorithm is a way to tell the value $<C_i(P,K),\Gamma_H>$, $i\in\{1,2\}$ from the consumption measurement at the selected sample. That is why, to
simplify this attack, we only select the output mask ($\Gamma_H$) to be $0x10$ or $0x20$ because then, we just have to separate the traces in two, the ones that have power measures greater
than the average power measure $S_1$ and the others $S_0$, assuming that the power traces in $S_1$ are such that $<C_i(P,K),0x20> = 1$ and $<C_i(P,K), 0x20> = 0$ for the others. We then 
assume that the power traces in $S_1$ are such that $<C_i(P,K),0x10> = 0$ and $<C_i(P,K), 0x10> = 1$ for the others since there is very few chance to have $C_i(P,K) < 0x10$ or $C_i(P,K) \geq 0x30$
from random plaintexts.
\begin{table}[!h]
\tiny{
\begin{center}
\begin{tabular}{|c|c|l|}
\hline
$\Gamma_H$ & Bias & Equation \\
\hline
\hline
0x10 & 0.0219 & 0 + P[5]+ P[26]+ P[27]+ P[31]+ P[45]+ P[53]+ P[61]+ K[6]+ K[7]+ K[29]+ K[38]+ K[52] \\
0x20 & 0.0215 & 1 + P[5]+ P[26]+ P[27]+ P[31]+ P[45]+ P[53]+ P[61]+ K[6]+ K[7]+ K[29]+ K[38]+ K[52] \\
0x20 & 0.0134 & 0 + P[28]+ P[29]+ P[31]+ P[37]+ P[45]+ P[53]+ K[6]+ K[7]+ K[29]+ K[61] \\
0x20 & 0.0156 & 1 + P[5]+ P[28]+ P[29]+ P[31]+ P[37]+ P[45]+ K[6]+ K[29]+ K[38]+ K[61] \\
0x10 & 0.0142 & 0 + P[5]+ P[28]+ P[29]+ P[31]+ P[37]+ P[45]+ K[6]+ K[29]+ K[38]+ K[61] \\
0x20 & 0.0189 & 1 + P[5]+ P[28]+ P[29]+ P[31]+ P[37]+ P[53]+ K[7]+ K[29]+ K[38]+ K[61] \\
0x10 & 0.0189 & 0 + P[5]+ P[28]+ P[29]+ P[31]+ P[37]+ P[53]+ K[7]+ K[29]+ K[38]+ K[61] \\
0x10 & 0.0126 & 1 + P[26]+ P[27]+ P[37]+ P[45]+ P[53]+ P[61]+ K[6]+ K[7]+ K[52]+ K[61] \\
0x20 & 0.0163 & 0 + P[5]+ P[8]+ P[9]+ P[37]+ P[45]+ P[53]+ P[61]+ K[6]+ K[7]+ K[38]+ K[52]+ K[61] \\
0x10 & 0.0167 & 1 + P[5]+ P[8]+ P[9]+ P[37]+ P[45]+ P[53]+ P[61]+ K[6]+ K[7]+ K[38]+ K[52]+ K[61] \\
0x10 & 0.0215 & 1 + P[5]+ P[14]+ P[15]+ P[31]+ P[37]+ P[45]+ P[61]+ K[6]+ K[29]+ K[38]+ K[52]+ K[61] \\
0x10 & 0.0146 & 0 + P[5]+ P[28]+ P[29]+ P[31]+ P[37]+ P[45]+ P[61]+ K[6]+ K[29]+ K[38]+ K[52]+ K[61] \\
0x10 & 0.0148 & 1 + P[5]+ P[8]+ P[9]+ P[31]+ P[37]+ P[45]+ P[61]+ K[6]+ K[29]+ K[38]+ K[52]+ K[61] \\
0x20 & 0.0223 & 0 + P[5]+ P[14]+ P[15]+ P[31]+ P[37]+ P[45]+ P[61]+ K[6]+ K[29]+ K[38]+ K[52]+ K[61] \\
0x20 & 0.0182 & 0 + P[5]+ P[28]+ P[29]+ P[31]+ P[37]+ P[53]+ P[61]+ K[7]+ K[29]+ K[38]+ K[52]+ K[61] \\
0x10 & 0.0152 & 0 + P[5]+ P[26]+ P[27]+ P[31]+ P[37]+ P[53]+ P[61]+ K[7]+ K[29]+ K[38]+ K[52]+ K[61] \\
0x10 & 0.0187 & 1 + P[5]+ P[28]+ P[29]+ P[31]+ P[37]+ P[53]+ P[61]+ K[7]+ K[29]+ K[38]+ K[52]+ K[61] \\
0x20 & 0.0157 & 1 + P[5]+ P[26]+ P[27]+ P[31]+ P[37]+ P[53]+ P[61]+ K[7]+ K[29]+ K[38]+ K[52]+ K[61] \\
0x20 & 0.0191 & 0 + P[5]+ P[26]+ P[27]+ P[31]+ P[37]+ P[45]+ P[53]+ P[61]+ K[6]+ K[7]+ K[29]+ K[38]+ K[52]+ K[61] \\
0x10 & 0.0183 & 1 + P[5]+ P[26]+ P[27]+ P[31]+ P[37]+ P[45]+ P[53]+ P[61]+ K[6]+ K[7]+ K[29]+ K[38]+ K[52]+ K[61] \\
\hline
\end{tabular}
\caption{Attack on DPA-contest traces Results}
\label{table:equations}
\end{center}
}
\end{table}

With this setup, and only considering these 11 equations, the 6 keys bits are retrieved from the first 2000 traces.

\end{document}